\documentclass[prc,showpacs,floatfix,preprint]{revtex4-1}
\usepackage{amsmath}
\usepackage{epsfig}
\usepackage{color}
\usepackage{textcomp}
\newcommand{\re}{\ref}
\newcommand{\be}{\begin{equation}}
\newcommand{\ee}{\end{equation}}

\newcommand{\la}{\label}
\newcommand{\ber}{\begin{eqnarray}}
\newcommand{\eer}{\end{eqnarray}}

\begin{document}

\title{Calculation of the astrophysical $S$-factor $S_{12}$ with the Lorentz integral transform}

\author{ Sergio Deflorian$^{1,2}$,  Victor D. Efros$^{3,4}$  and Winfried 
Leidemann$^{1,2}$ 
  }

\affiliation{
  $^{1}$Dipartimento di Fisica, Universit\`a di Trento, I-38123 Trento, Italy \\
  $^{2}$Istituto Nazionale di Fisica Nucleare, TIFPA,
  I-38123 Trento, Italy \\
  $^{3}$ National Research Centre "Kurchatov Institute", 123182 Moscow, Russia\\
  $^{4}$ National Research Nuclear University MEPhI(Moscow Engineering Physics Institute),
  Moscow, Russia   
}

\begin{abstract}
The LIT approach is tested for the calculation of astrophysical $S$-factors.
As an example the $S$-factor of the reaction $^2$H($p,\gamma)^3$He is considered.
It is discussed that a sufficiently high density of LIT states at low energies is necessary
for a precise determination of $S$-factors. In particular it is shown that the
hyperspherical basis is not very well suited for such a calculation and that a
different basis system is much more advantageous. A comparison of LIT results with
calculations, where continuum wave functions are explicitly used, shows that the LIT approach
leads to reliable results. It is also shown how an error estimate of the LIT inversion
can be obtained.
\end{abstract}

\maketitle

\section{Introduction}

The study of stellar nucleosynthesis is one of the central issues of nuclear astrophysics.
In order to understand the details of this process it is necessary to
have a precise determination of a large number of reaction cross sections
at relatively low energies. Considering for example the solar proton-proton cycle
and taking into account that the temperature of the core of the sun is about 1.5$\times$10$^7$ K 
one finds that the relevant energies are below 100 keV~\cite{AdG11}. At such low energies cross sections can become extremely
small, in particular in presence of a Coulomb barrier between the reacting
particles. In many cases data have been obtained only at higher energies, 
which makes extrapolations to lower energies necessary. Therefore it is very
helpful to have additional input from the theory side, especially calculations
with ab initio methods~\cite{LeO13,CaD14} employing modern realistic nuclear forces can help to 
reduce error estimates for cross sections. 

Among the relevant nuclear reactions of astrophysical interest there are many electroweak processes.
Concerning such kind of reactions the Lorentz integral transform (LIT)~\cite{EfL94} is a particularly 
interesting ab initio method, since it reduces a continuum-state problem to a much simpler to solve bound-state 
like problem, however, involves an inversion of the transform~\cite{BaE10,AnL05,Lei08}. In the past the LIT 
was applied to quite a number 
of reactions~\cite{EfL07,BaS14}, where in most cases 
the bound-state methods of choice were expansions in hyperspherical harmonics (HH). 
Up to today the LIT was never applied to calculations of  cross sections
relevant in stellar nucleosynthesis. In fact extremely small low-energy cross sections are 
a challenge for the method because of the above mentioned LIT inversion. In such a scenario one needs 
a rather high density of LIT states in the low-energy region in order to have a sufficient resolution of the LIT. 
That such a request can be problematic became evident in recent LIT calculations
for the $^4$He isoscalar monopole resonance~\cite{BaB13}, where the 
effective interaction HH expansion technique~\cite{BaL00,BaE04} was applied. On the one hand the resonance strength
was successfully determined, on the other hand the resonance width could not be computed since the density of LIT states
was much too low in the resonance region. In~\cite{Lei15} it was then shown that with a four-body hybrid basis,
consisting of a three-body HH basis plus a single-particle basis, one obtains a much higher density of LIT states
in the $^4$He isoscalar monopole resonance region, which is located below the three-body breakup
threshold.   

The aim of the present paper is to check whether the LIT method succeeds to reliably determine the low-energy
cross section in presence of a Coulomb barrier. To this end we have chosen to calculate the $S$-factor $S_{12}$
of the reaction $^2$H$(p,\gamma)^3$He. A positive outcome of the check would allow to apply the LIT method
also for the calculation of $S$-factors involving a higher number of nucleons. The calculation is carried out in two different ways:
(i) via the LIT method and (ii) with the explicit calculation of the $d$-$p$ continuum wave function.
For this check it is not necessary to use a realistic nuclear force, therefore we take the central 
MT-I/III potential~\cite{MaT69}
as $NN$ interaction, however we would like to mention that $S_{12}$ was calculated in rather complete ab initio
calculations~\cite{MaV05,MaM16}.

The paper is organized as follows. After the definition of the $S$-factor $S_{12}$ in section~II, 
in subsection II-A the LIT approach for the calculation
of the $S$-factor is described. Since we want to determine the $S$-factor also in
the conventional way, in subsection II-B we discuss the calculation of continuum states 
with the Kohn variational principle. Section III contains a detailed study of the LIT method.
It is shown that the density of LIT states in the low-energy region depends significantly
on the basis system chosen for the solution of the LIT equation. The section closes with
a comparison of LIT and conventional results for the low-energy $^3$He photodisintegration cross 
section and $S$-factor $S_{12}$ and with a brief summary.

\section{Calculation of the $S$-factor $S_{12}$}

The $S$-factor $S_{12}$ is defined as follows
\begin{equation}
\label{S-fac}
 S_{12}(E) = \sigma_{\rm cap} \, E \,\exp(2\pi\eta) \,,
\end{equation}
where $\sigma_{\rm cap}$ is the cross section of the reaction $d+p \rightarrow \,^3$He$+\gamma$,
$E$ denotes the relative energy of the deuteron-proton pair, and $\exp(2\pi\eta)$ is the Gamow factor
taking into account the effect of the Coulomb barrier with 
\begin{equation}
 \eta = \sqrt{{\frac{\mu c^2}{2E}}} \,\alpha \,,
\end{equation}
where $\mu$ is the reduced mass of the deuteron-proton pair and $\alpha$ is the fine structure constant.
 
We determine $\sigma_{\rm cap}$
by first calculating the cross section $\sigma_\gamma$ of the inverse reaction
$^3$He$+\gamma \rightarrow d+p$ and then using the relation
\begin{equation}
\label{cap}
\sigma_{\rm cap}(E) = {\frac{2 E_{\gamma}^2}{3 k^2}} \sigma_\gamma(E_\gamma) \,,
\end{equation}
where $E_{\gamma}$ is the photon energy and $k$ denotes the relative momentum of the
deuteron-proton pair. The photodisintegration cross section of $^3$He is calculated
in unretarded dipole approximation,
\begin{equation}
\label{xsec}
\sigma_\gamma(E_\gamma) = 4 \pi^2 \alpha E_\gamma R(E_\gamma) \,,
\end{equation}
where
\begin{equation}
\label{response}
R(E_\gamma) = \int df |\langle f| D_z | 0\rangle|^2 \delta(E_f - E_0 - E_\gamma) 
\end{equation}
is the dipole response function. In Eq.~(\ref{response}) $|0 \rangle$ and $|f\rangle$ are the $^3$He ground state 
and the deuteron-proton final state, respectively, while $E_0$ and $E_f$ are the 
corresponding eigenenergies. Finally,
$D_z$ is the third component of the nuclear dipole operator.

As mentioned in the introduction we calculate $ R(E_\gamma)$ in two different ways: 
(i) with the LIT approach, where bound-state methods
can be used, and (ii) with the explicit calculation of the continuum state $|f\rangle$.
Both methods are described briefly in the following two subsections.

\subsection{Calculation with LIT approach}
The LIT of the response function $R(E_\gamma)$ is defined as follows
\begin{equation}
\label{LIT}
L(\sigma) = \int dE_\gamma \, {\cal L}(E_\gamma,\sigma) \, R(E_\gamma) \,, 
\end{equation}
where the kernel ${\cal L}$ is a Lorentzian with a width of $2\sigma_I$, which is located at $E_\gamma=\sigma_R$:
\begin{equation}
{\cal L}(E_\gamma,\sigma  = \sigma_R + i \sigma_I) = {\frac {1}{(E_\gamma-\sigma_R)^2 + \sigma_I^2}} \,.
\end{equation}
In fact the width can in principle
be adjusted to resolve the detailed structure of $R(E_\gamma)$ and
due to the variable width the LIT is a transform with a controlled resolution.  However, an increase of the resolution by a reduction 
of $\sigma_I$ does not come for free and it requires in general an increase of the precision of the calculation. 

The LIT $L(\sigma)$  is calculated by solving the following equation
\begin{equation} 
\label{eqLIT}
(H-E_0-\sigma) \, |\tilde\Psi(\sigma) =  D_z | 0\rangle \,,
\end{equation}
where $H$ is the Hamiltonian of the particle system under consideration. The solution $\tilde\Psi(\sigma)$ is localized, 
since the rhs of Eq.(\ref{eqLIT}) is asymptotically vanishing.
Therefore one can determine $\tilde\Psi(\sigma)$ using bound-state methods. 
The solution directly leads to the transform:
\begin{equation}
L(\sigma) = \langle \tilde\Psi(\sigma) | \tilde\Psi(\sigma) \rangle \,.
\end{equation}
Finally, the response function $R(E_\gamma)$ is obtained from the inversion of
the transform (for details see \cite{EfL07,Lei08,AnL05,BaE10}).

Here we solve the LIT equation~(\ref{eqLIT}) via an expansion on a complete basis, where the number of basis functions
$N$ is increased up to a sufficient convergence. One can understand such an expansion for the solution of the LIT equation as follows.
The spectrum of the Hamiltonian for the basis is determined, thus one has $N$ eigenstates $\phi_n$ with eigenenergies $E_n$. 
The LIT solution assigns to any eigenenergy $E_n$ a LIT state, which is a Lorentzian with strength $S_n$ and width $2\sigma_I$. 
The strength $S_n$ depends on the source term on the rhs of the LIT equation: 
\begin{equation}
S_n = |\langle \phi_n| D_z | 0 \rangle |^2 \,. 
\end{equation}
The LIT result is then just given by the the sum over the $N$ LIT states: 
\begin{equation}
\label{LIT_En}
 L(\sigma) = \sum_{n=1}^N {\frac {S_n}{(\sigma_R-(E_n-E_0))^2 + \sigma_I^2}} \,.
\end{equation}
From the equation above it is evident that at a given resolution of the LIT,
which is characterized by the value of $\sigma_I$, one needs a sufficient density
of LIT states as discussed in detail in Ref.~\cite{Lei15}.
There it is illustrated that the density of LIT states is not only correlated
to the number of basis functions $N$, but depends also on the specific basis.
For example, for the electromagnetic $^4$He breakup it was discussed that it is
very difficult to increase the density of LIT states below the three-body breakup
for a hyperspherical harmonics (HH) basis. As is discussed in the following section 
a similar problems occurs at use of the HH basis also in the three-body case considered in
the present work. At this point we would like to emphasize that 
the LIT contains in general for a generic electroweak reaction the full response function $R$
with all breakup channels and 
one may use any complete localized $A$-body basis set for the calculation of the LIT. 
On the other hand one has to have in mind that in a given energy range
one basis set can be more advantageous than another one.

In order to take into account the findings of~\cite{Lei15} we use for the LIT calculation
two different basis systems. A HH basis with two-body correlations of the Jastrow type
as was done for the same $NN$ potential in~\cite{EfL97}. For the second basis we use
the two Jacobi coordinates of the three-body system in an explicit way, therefore this basis
will be called Jacobi basis. The spatial part of this basis starts from the following
definition
\begin{eqnarray}
\psi_{n_1,n_2,\,l_1,\,l_2} & = & \sum_{m_1m_2}\mathcal R^{[1]}_{n_1}\left(\eta_1\right)Y_{l_1}^{m_1}\left(\theta_1,\phi_1\right)\nonumber\\
& & \times \mathcal R^{[2]}_{n_2}\left(\eta_2\right)Y_{l_2}^{m_2}\left(\theta_2,\phi_2
\right) \langle L=1\,M|l_1m_1l_2m_2 \rangle \label{spatial} \,,
\end{eqnarray}
where $\boldsymbol\eta_1 =(\eta_1,\theta_1,\phi_1)$ is the relative ("pair") coordinate of particles 1 and 2, 
$\boldsymbol\eta_2 =(\eta_2,\theta_2,\phi_2)$ is the single-particle coordinate of the third particle with respect
to the center of mass of particles 1 and 2,
the $Y_{l}^{m}(\theta,\phi)$ are spherical harmonics and $\langle L=1\,M|l_1m_1l_2m_2 \rangle$ denotes
a Clebsch-Gordan coefficient (note that because of the dipole operator in Eq.~(\ref{response}) one
needs only basis states with angular momentum $L=1$).
The radial functions $\mathcal{R}^{[1,2]}_{n}
(\eta)$ are defined as follows
\begin{eqnarray}\label{lag1}
\mathcal R^{[1]}_{n_1}(\eta_1) & = & \sqrt{\frac{n_1!}{(n_1+2)!}}L_{n_1}^{(2)}(\frac{\eta_1}{b_1})e^{-\frac{\eta_1}{2b_1}}
b_1^{-\frac{3}{2}}\\
\mathcal R^{[2]}_{n_2}(\eta_2) & = & \sqrt{\frac{n_2!}{(n_2+2)!}}L_{n_2}^{(2)}(\frac{\eta_2}{b_2})e^{-\frac{\eta_2}{2b_2}}
b_2^{-\frac{3}{2}}\ \ ,\label{lag2}
\end{eqnarray}
where $L_{n_i}^{(2)}$ is a Laguerre polynomial of order $n_i$ ($n_i \in \{0,1,2, ...,N_i-1\}$) with parameter $b_i$.
This is very similar to our expansions of the HH hyperradial function $R_n$, in fact, in this case we have
\begin{equation}
\label{hyperradial}
\mathcal R_n(\rho) = \sqrt{\frac{n!}{(n+5)!}}L_{n}^{(5)}(\frac{\rho}{b})e^{-\frac{\rho}{2b}}
b^{-\frac{3}{2}} \,,
\end{equation} 
where $\rho$ is the hyperradius and $n \in \{0,1,2, ...,N-1\}$.

Including the spin-isospin part to $\psi_{n_1,n_2,\,l_1,\,l_2}$ of Eq.(\ref{spatial}) one has
\begin{equation}
\label{phi}
\phi_{n_1,n_2,\,l_1,\,l_2,s_{12},t_{12}}=\psi_{n_1,n_2,\,l_1,\,l_2}\chi^S(s_{12})\chi^T(t_{12})\ \ ,
\end{equation}
where the spin and isospin functions $\chi^S(s_{12})$ and $\chi^T(t_{12})$ are defined to have spin $s_{12}$ 
and isospin $t_{12}$ equal to 1 or 0 for the first two
particles and total spin and isospin $S=\frac{1}{2}$ and $T=\frac{1}{2}$. 
A totally antisymmetric basis state is given by 
\begin{equation}
\Phi_{n_1,n_2,\,l_1,\,l_2,s_{12},t_{12}}={\cal A}\, \phi_{n_1,n_2,\,l_1,\,l_2,s_{12},t_{12}},
\end{equation}
where ${\cal A}$ is a proper antisymmetrization operator.

\subsection{Explicit calculation of the continuum states}

To obtain the deuteron--proton final states entering Eq.~(\re{response}) we apply the 
version~\cite{scat} of the
general trial function approach which employs the HH expansion.  
The continuum wave function is written as 
$\Psi_f=X+Y$ where at large distances 
the $Y$ component represents the two--body asymptotics of $\Psi_f$. 
The $X$ component is an expansion over HH. At energies below 
the three--body breakup threshold it 
vanishes at large distances  and above the threshold
it reproduces the three--body breakup asymptotics in the absence of the Coulomb interaction. 

Our calculation refers to the former case. One sets
\be X=\sum_{i=1}^{i_{max}}c_i\psi_i\ee
where $\psi_i$ are basis functions. They are the sums that are
antisymmetric with respect to nucleon permutations 
of products of correlated
hyperspherical harmonics mentioned above
and spin--isospin functions, 
 times the Laguerre type hyperradial basis functions (\re{hyperradial}). 
The $c_i$ expansion coefficients are to be determined.

The $Y$ component     
is of the form $Y_R+Y_I\tan\delta$ where $\delta$ is the 
trial scattering phase shift. The functions $Y_{R,I}$ 
are of the form \mbox{${\cal A}\varphi(12,3)$}
where 1, 2, and 3 
are the nucleon numbers, and ${\cal A}$ is the antisymmetrization operator.
The function $\varphi(12,3)$ is the product of a channel function with a given
spin and isospin of a system 
and a relative motion function pertaining to a given orbital momentum $L$. 
The channel function is
obtained
by coupling the deuteron wave function of the nucleons 1 and 2 and 
the spin--isospin function of the nucleon 3. The relative motion function
is the product of the spherical harmonics and a radial function.

The radial function is $(kr)^{-1}F_L(kr,\eta)$ in the case of $Y_R$, and
$g_L(r)(kr)^{-1}G_L(kr,\eta)$ in the case of $Y_I$. 
Here 
$F_L$ and $G_L$ are the regular and irregular Coulomb functions, and 
$g_L(r)$ is a correction factor. It is to be taken 
such that  $g_L(r)$ turns to unity
beyond the interaction region and $g_LG_L$ is regular 
and behaves  e.g. like $F_L$ at $r\rightarrow0$. In our $L=1$ case we used  
$g_1(r)=[1-\exp(-r/r_0)]^3$, $r_0$ being a scale parameter, which is of the same form 
as in~\cite{MaV05}. The results vary little in a broad range
of $r_0$ values when convergence is achieved.

The above trial wave function may be written as
\be \Psi_f=\sum_{i=0}^{i_{max}}c_i\psi_i+Y_R\ee
where $\psi_0=Y_I$ and $c_0=\tan\delta$. The system of equations 
\be \sum_{i=0}^{i_{max}}\langle\psi_j|H-E|\psi_i\rangle c_i=
-\langle\psi_j|(H-E)Y_R\rangle\la{eqs}\ee
with $j=0,\ldots,i_{max}$ was used to obtain the $c_i$ coefficients. 
These equations emerge in particular from the requirement for the 
Kohn functional
to be stationary.

At a given $i_{max}$ value, the quality of the wave function thus obtained  
apparently deteriorates when
the energy approaches the eigenvalues of the $\langle\psi_j|H|\psi_i\rangle$ matrix. Corresponding vicinities
  of the 
eigenvalues in which results are unsatisfactory  normally are narrow as compared  
to distances between the eigenvalues~\cite{schwartz}. The least--square method 
involving in addition to Eqs.~(\re{eqs})  the equations of the same form with $j$ exceeding $i_{max}$ may cure the 
deficiency~\cite{schmid}. In our 
low--energy case, Eqs.~(\re{eqs}) did not lead to problems in the range of $i_{max}$ considered so that the
convergence trends of the results do not seem to depend on energy. 

Let us denote $\delta\Psi_f$ the deviation  of the approximate 
$\Psi_f$  wave function from the exact one. 
The difference between the  exact $\tan\delta$ and its  value
that pertains to an approximate $\Psi_f$  
may be represented as an integral with the integrand containing $\delta\Psi_f$ linearly. 
In the difference to this, the deviation of the Kohn functional value 
from the exact $\tan\delta$ is quadratic
with respect to $\delta\Psi_f$. Thus the value of the Kohn functional is more accurate 
than obtained directly from Eqs. (\re{eqs}) when 
a calculation is close to convergence.
We replaced $c_0\equiv\tan\delta$ with the value of the Kohn functional 
in the  equations~(\re{eqs})  with  $j\ge1$ 
to get  the rest $c_1,\ldots,c_{i_{max}}$ coefficients such that the equations are satisfied.
However, these coefficients 
are not necessarily more accurate than those obtained directly from Eqs.~(\re{eqs}).

For checking purposes we  compared our $P$ wave phase shifts   with those  
obtained with the same MT-I/III potential
by the Pisa group~\cite{Kie16}. Their scattering calculations are known 
to be of a high precision~\cite{DeF05}. The differences found
between the Kohn functional values of the phase shifts
are about 0.5\% or less \cite{Def16}. 

\section{Discussion of results}
\vskip 0.5cm
\begin{figure}[htb]
\centerline{\includegraphics[width=0.9\columnwidth]{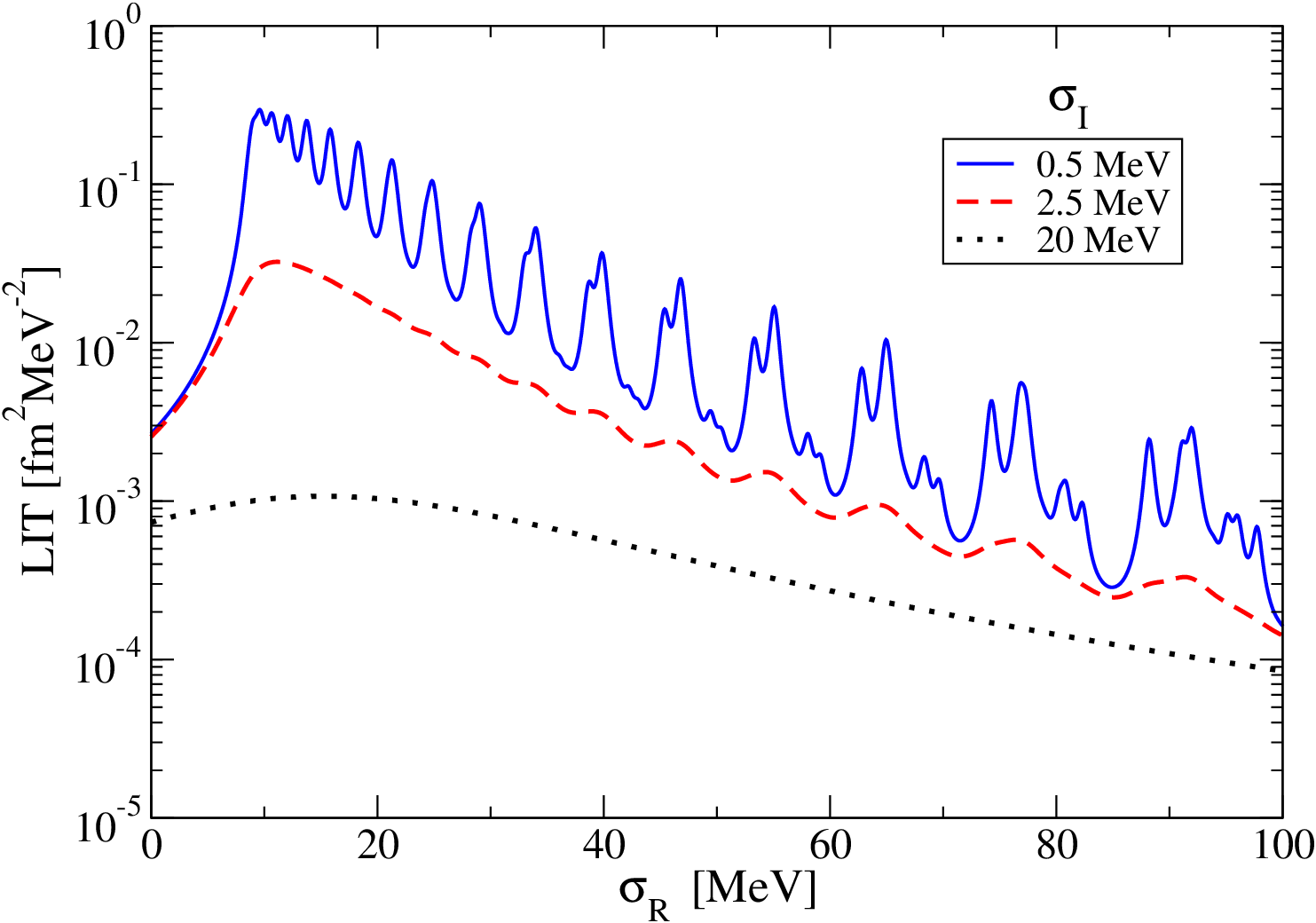}}
\caption{LIT of the $^3$He dipole response function for $T=1/2$ with $\sigma_I=0.5$, 2.5 and 20 MeV
calculated with an HH basis of 30 hyperspherical and 31 hyperradial states for a total of 930 basis
states ($b=0.3$ fm).
}
\end{figure}
We start the discussion illustrating first results, where the HH basis is used 
for the calculation of the LIT of the $^3$He photodisintegration. We consider 
only the final state in the isospin $T=1/2$ channel, since the $T=3/2$ channel 
corresponds exclusively to a three-body breakup. In Fig.~1 we show results for 
various values of $\sigma_I$. One sees that with $\sigma_I=20$ MeV a smooth 
transform is obtained, then with an increase of the resolution to $\sigma_I=2.5$ 
MeV the transform starts to have an oscillating behaviour beyond 20 MeV, and a 
still further increase of the resolution to $\sigma_I=0.5$ MeV exhibits the 
underlying structure of the single LIT states (see Eq.~(\ref{LIT_En})). From 
the last result one can conclude that the resolving power of the LIT is certainly 
not just given by the chosen $\sigma_I$ value. 

For a higher degree of resolution one has to increase the density of LIT states,
which can be achieved in two ways, namely by increasing the number of basis functions and by enhancing the $b$ parameter
of the hyperradial wave function of Eq.~(\ref{hyperradial}). Both measures are taken for the results shown in Fig.~2, where
we illustrate the low-energy part of the LIT for rather small $\sigma_I$ values. It is evident that the
density of LIT states grows as expected. In Fig.~2d one observes a rather high LIT state density and one could
easily further increase the density. However, one readily sees that there is
not a single LIT state below the three-body breakup threshold at about 8 MeV ($^3$He binding energy with
MT-I/III potential). On the other hand the calculation of the $S$-factor $S_{12}$ requests energies just beyond
the two-body breakup threshold at 5.8 MeV (difference of binding energies of $^2$H and $^3$He for the MT-I/III potential). 
Thus one cannot expect that an inversion of the LITs of Fig.~2 leads
to a high-precision result in the region of astrophysical relevance. Here a comment is in order concerning
the use of a more realistic nuclear force for a calculation of the LIT with the HH basis. In this case one can find 
a few LIT states below the three-body breakup threshold, but also there 
one encounters the problem of further increasing the density of LIT states in a systematic way
in order to obtain a smooth LIT with a sufficiently small $\sigma_I$~\cite{BaB13,Lei15}.
\vskip 1cm
\begin{figure}[htb]
\centerline{\includegraphics[width=0.9\columnwidth]{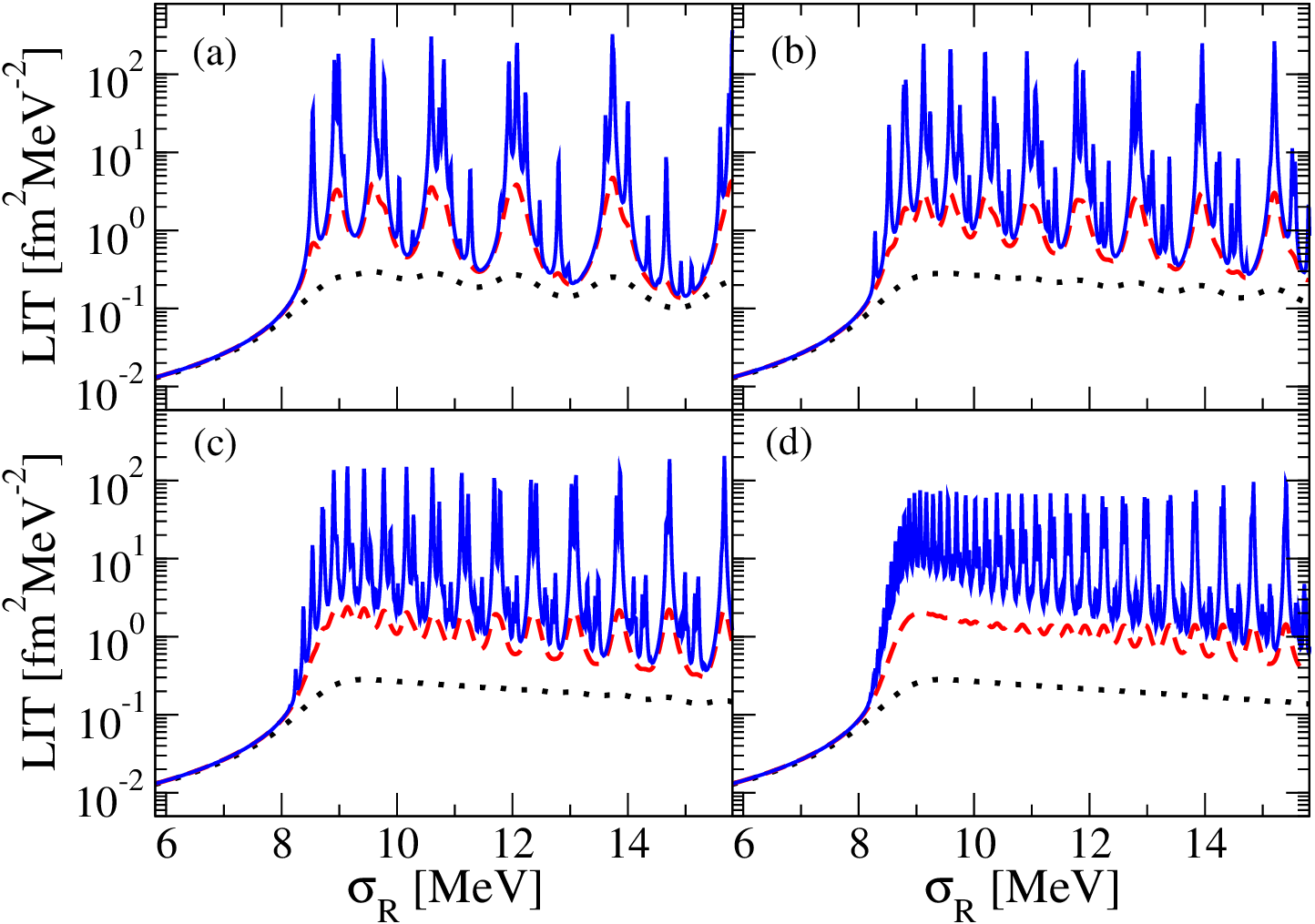}}
\caption{Low-energy part of the LIT of Fig.~1 with $\sigma_I=0.01$ (full), 0.1 (dashed) and 0.5 MeV (dotted),
results with different HH basis systems: (a) same basis as in Fig.~1,
(b) 40 hyperspherical and 51 hyperradial states ($b=0.3$ fm),
(c) as in (b), but with $b = 0.5$ fm, 
(d) 40 hyperspherical and 76 hyperradial states ($b = 1$ fm).
}
\end{figure}
Now we turn to the results with the Jacobi basis. Since in principle we are only interested in 
the cross section just above the two-body breakup threshold we only consider $S$-wave interaction
for the pair coordinate, this means that in Eq.~(\ref{spatial}) only basis states with
$l_1=0$ and $l_2=1$ are taken into account (as already mentioned for the dipole response we have $L=1$).
For the radial parts of the pair and single-particle wave functions we choose $b_1=0.75$ fm
and $b_2=0.5$ fm, respectively.
\vskip 1.5cm
\begin{figure}[htb]
\centerline{\includegraphics[width=0.7\columnwidth]{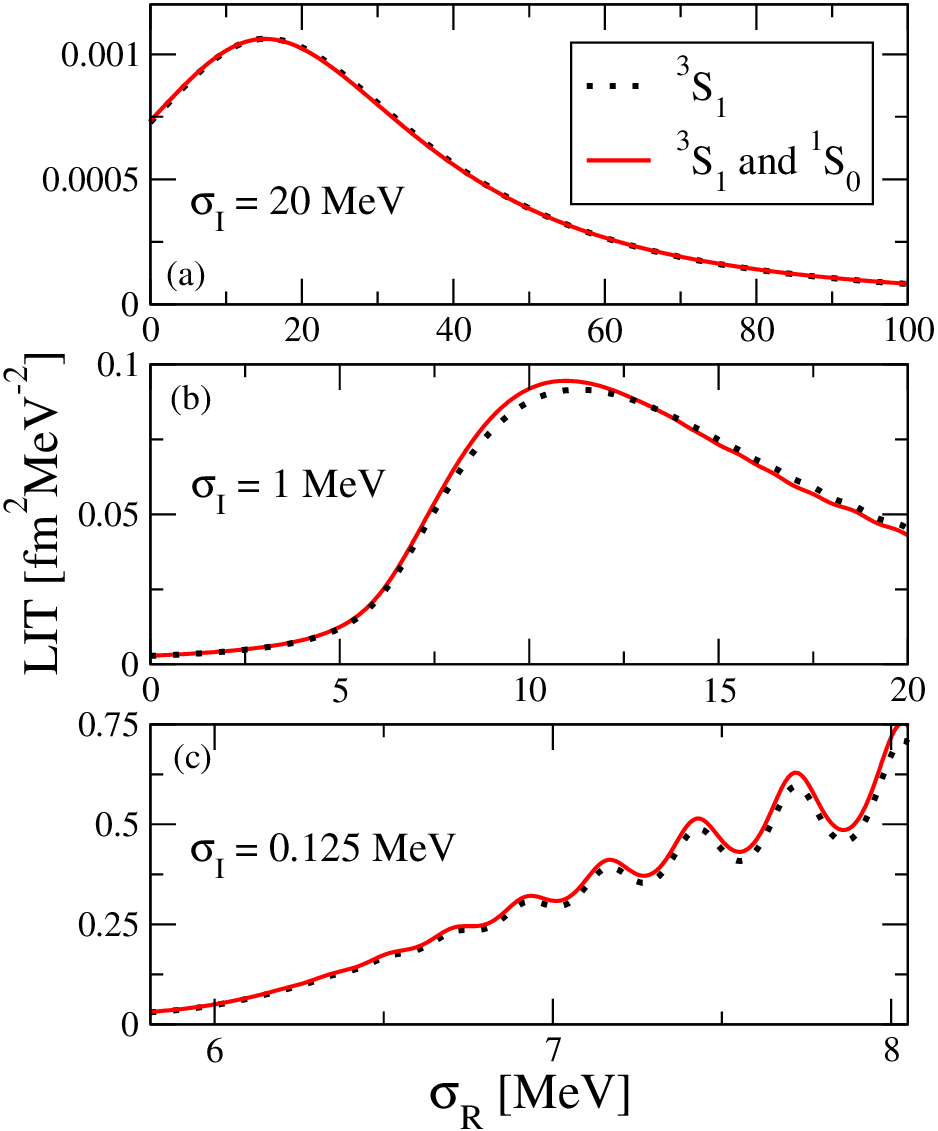}}
\caption{LIT of the $^3$He dipole response function for $T=1/2$ with $\sigma_I=20$, 1 and 0.125 MeV
calculated with the Jacobi basis taking into account only $^3S_1$ states in the pair coordinate of $\phi$
(dotted) and in addition also $^1S_0$-states (full).
}
\end{figure}
In Fig.~3 we show the LIT for the cases
that the pair in $\phi$ of Eq.~(\ref{phi}) is solely in a $^3S_1$-state and the additional effect when also $^1S_0$-states are allowed.
One sees that the contribution due to the $^1S_0$-states is quite tiny. 
In fact a rather small number of basis states with the pair in the $^1S_0$ state 
($N_1 = 5$, $N_2 = 19)$ is sufficient in order to obtain convergence. As shown in Fig.~4
the convergence of the main LIT contribution due to the $^3S_1$-states is not as rapid as in case of the $^1S_0$ states. 
On the one hand one needs only a rather moderate value for $N_1$ of about 20 to obtain a sufficient convergence 
in the pair coordinate as shown in Fig.~4b (note a result with $N_1$=24 could not be distinguished in
the figure from the $N_1 = 19$ result). On the other hand the situation is different for 
the single-particle coordinate (see Fig.~4a). In order to have a sufficiently convergent LIT in the region just above the 
two-body breakup threshold with a small $\sigma_I$ value of 0.125 MeV one has to go up to an $N_2$ of about 70.
In fact for our calculation of the $S$-factor $S_{12}$ we use $N_2=79$. 

It is interesting to observe the different effects of an increase of $N_1$ and $N_2$.
The enhancement of basis states for the pair coordinate in Fig.~4b shifts the transform to
lower energies without changing the shape of the LIT. This corresponds to an energy shift of the low-energy LIT states
to lower energies without a notable change of the density. On the contrary the increase of basis states for 
the single-particle coordinate (Fig.~4a) leads to a smoother result
of the transform due to an increased density of LIT states.
\vskip 2cm
\begin{figure}[htb]
\centerline{\includegraphics[width=0.8\columnwidth]{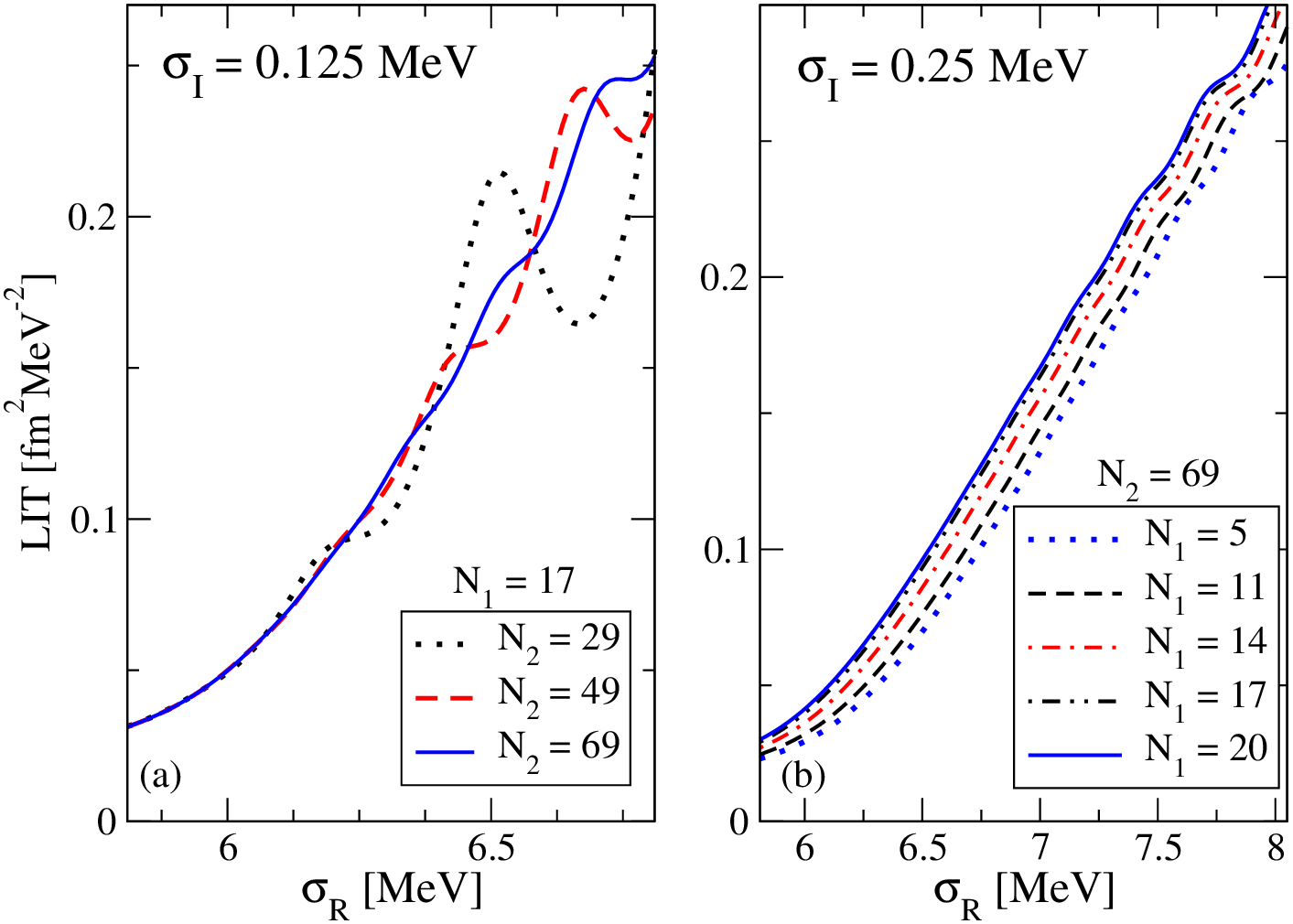}}
\caption{Convergence pattern for the LIT of the $^3$He dipole response function for $T=1/2$ calculated with the Jacobi basis:
(a) $N_1=17$ and various $N_2$ values ($\sigma_I=0.125$ MeV) and 
(b)  $N_2=69$ and various $N_1$ values ($\sigma_I=0.25$ MeV). 
}
\end{figure}
In Fig.~5 we compare the low-energy LIT calculated with HH and Jacobi basis systems for various $\sigma_I$ values. 
Note that different from the case with the Jacobi basis, where only $S$-wave interaction is taken into account,
for the HH basis also interaction in higher partial waves is considered, however, the contribution of the
latter should be quite small. In fact for large
$\sigma_I$ (see Fig.~5a) one can hardly find any difference between both results. Even for $\sigma_I=5$ MeV, shown in Fig.~5b,
the results are rather similar, whereas a decrease of $\sigma_I$ to 0.5 MeV, also shown in Fig.~5b, exhibits quite some 
difference:
the peak of the LIT of the HH basis is considerably more pronounced than that of the Jacobi basis. 
To a large extent the difference is caused by the missing LIT states at low energy for the HH basis
and not by the additional interaction in higher partial waves. Thus one may conclude that the lack of low-energy 
LIT states leads to a shift
of low-energy strength to the peak region region just above the two-body breakup threshold. 

The energy distribution of low-energy
LIT states for both basis systems is nicely illustrated in Fig.~5c for $\sigma_I=0.01$ MeV. Only for the Jacobi basis
one finds LIT states directly above the two-body breakup threshold. The LIT state density is so high that one obtains a
smooth LIT in the very threshold region even with $\sigma_I=0.1$ MeV and up to the three-body breakup threshold with
$\sigma_I=0.25$ MeV.
\vskip 1cm
\begin{figure}[htb]
\centerline{\includegraphics[width=0.65\columnwidth]{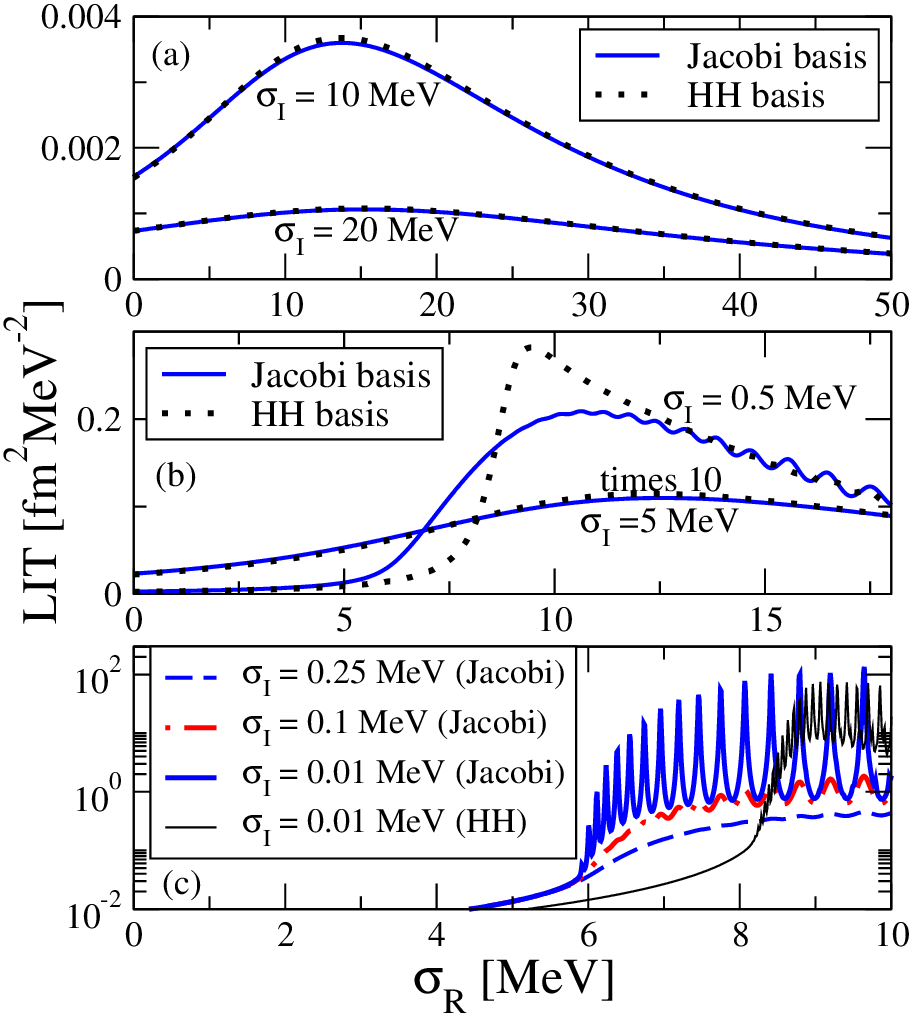}}
\caption{Comparison of the LIT calculated with HH and Jacobi basis systems with various $\sigma_I$ values as
indicated in the figure.
}
\end{figure}
In order to determine the cross section $\sigma_\gamma$ one has to invert the calculated transforms.
With regard to the aim to determine the $S$-factor it is evident that close to the threshold region one wants to work with a high resolution, however, one has to take
into account that with a small $\sigma_I$ value one does not obtain a smooth LIT at higher energies
because the density of LIT states decreases with growing energy. In fact it
is better to work with an energy dependent $\sigma_I$. Therefore we divide the $\sigma_R$ range in various
intervals [$E_j$, $E_{j+1}]$ ($j=1,2,3,...,J)$ and take in this interval $\sigma_I=\sigma_{I,j}$. Considering that
we have calculated the LIT for a certain number of $\sigma_R$ points $\sigma_{R,k}$ ($k=1,2,3,...,K)$ we rescale the LIT for all 
$\sigma_{R,k} \ge E_{j+1}$ by the factor 
\begin{equation}
\label{factor}
f(j+1)={\frac{L(\sigma_{k_2(j)},\sigma_{I,j})}{L(\sigma_{k_1(j+1)},\sigma_{I,j+1})}} \,, 
\end{equation}
where $\sigma_{k_1(j)}$ is the lowest
and $\sigma_{k_2(j)}$ the highest $\sigma_R$ value in interval  [$E_j$, $E_{j+1}]$. Note that this is made in a
cumulative way, thus for the LIT in the last interval ($\sigma_R \in [E_{J-1}$, $E_{J}]$) we have the total factor 
$F=f(2)f(3)...f(J)$. The values we have chosen for $E_j$ and $\sigma_{I,j}$ are given in Table~I. The application
of Eq.~(\ref{factor}) and the definitions given in Table~1 define a new transform ${\cal L}$.
\begin{table}
\caption{$E_j$ and $\sigma_{I,j}$ values for the definition of the new transform ${\cal L}$ 
}
\begin{center}
\begin{tabular}{c|c|c} \hline\hline
$j$ & $E_j$ [MeV] &  $\sigma_{I,j}$ [MeV]   \\
\hline
  1 & 5.7  & 0.125  \\
  2 & 6.15 & 0.175  \\
  3 & 6.55 & 0.35 \\
  4 & 8.05 & 0.7 \\
  5 & 10.55 & 1.1 \\
  6 & 13.05  & 1.5 \\
  7 & 23.05 & 5 \\
  8 & 58.05  & 10 \\
  9 & 108.05  &20 \\
  10 & 308.05 & - \\

\hline\hline
\end{tabular}
\end{center}
\end{table}

In Fig.~6 we show the newly defined transform ${\cal L}(\sigma)$, where we use
a Jacobi basis with ($N_1=24$, $N_2=79)$ and ($N_1=5$, $N_2=19)$ for the  $^3S_1$ states
and the $^1S_0$ states, respectively. The dashed curve in the figure shows the LITs $L$ for the various energy intervals
without any additional factor, whereas the continuous curve corresponds to the result when the additional
factors of Eq.~(\ref{factor}) are introduced. Note that according to the definition of the $f(j)$ the derivative
of ${\cal L}$ seems to be not continuous, but actually this is not the case since the transform is only defined pointwise
in $K$ $\sigma_R$ points. In principle one could also work with the transform described by the dashed curve in
Fig.~6, but this would mean that the impact of the transform is reduced with growing energy. The rescaling simulates the case where the transform is calculated with a single $\sigma_I$.
\vskip 1cm
\begin{figure}[htb]
\centerline{\includegraphics[width=0.75\columnwidth]{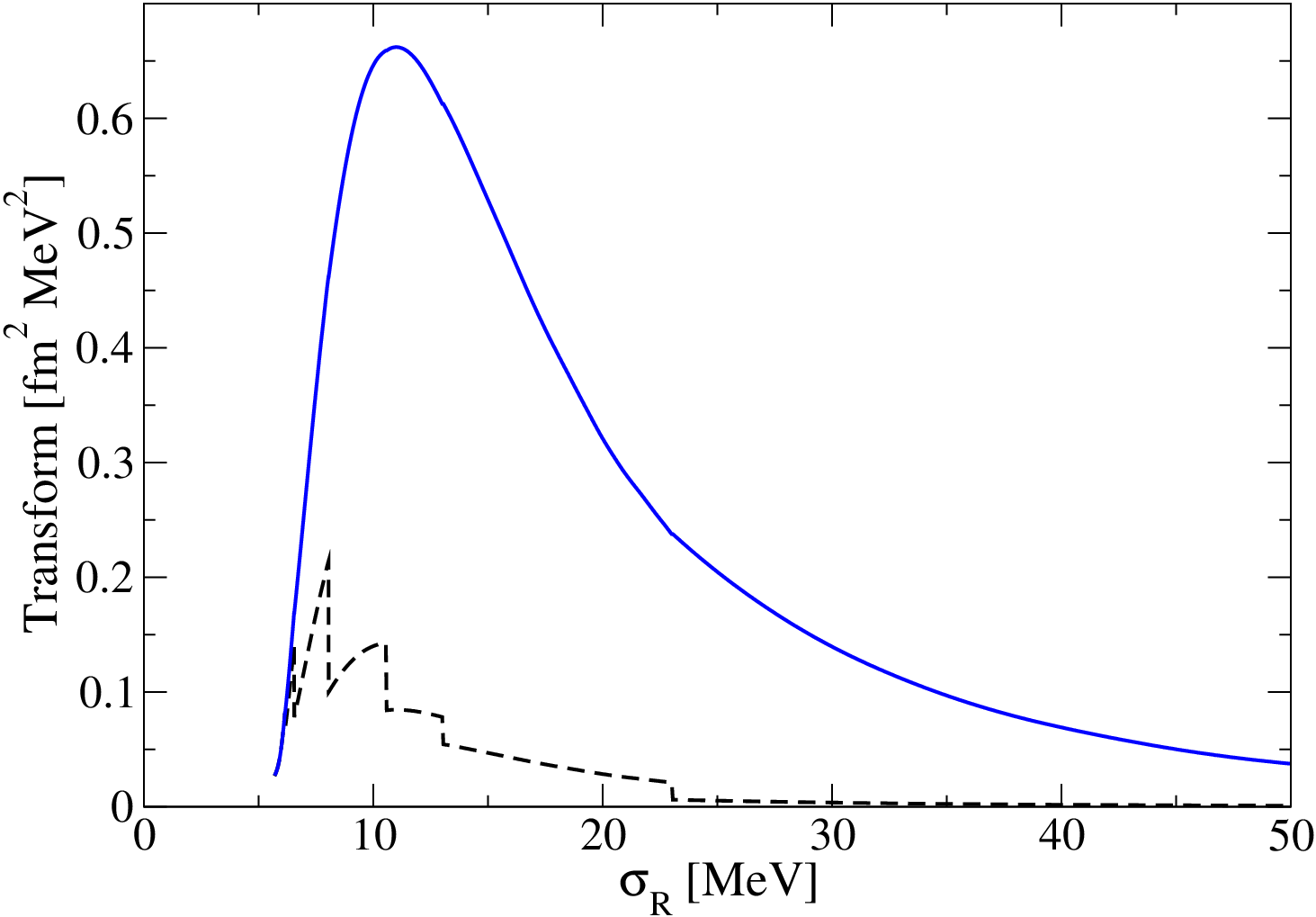}}
\caption{The LITs $L$ with $\sigma_I=\sigma_{I,j}$ in the various energy intervals $[E_j$, $E_{j+1}]$ (dashed)
and the new transform ${\cal L}$ (full) as described in the text.}
\end{figure}
For the inversion we use our standard method, where the response function $R$ is expanded as follows
\begin{equation}
\label{inv1}
R(E_\gamma=E+E_{thr}) = \sum_{n=1}^N c_n g_n(E) \,,
\end{equation}
where $E$ is defined as in Eq.~(\ref{S-fac}) and $E_{thr}$ is the energy of the two-body breakup threshold.
In order to consider the effect of the Coulomb barrier we include the Gamow factor of Eq.~(\ref{S-fac}) taking
\begin{equation}
\label{inv2}
g_n(E) = \exp(-2\pi\eta) \exp[(-\alpha E)/n] \,,
\end{equation}
where $\alpha$ is a non-linear parameter. The various $g_n(E)$ are then transformed numerically to 
the $\sigma$-space according to the LIT transformation given in Eq.~(\ref{LIT}) for the response function.
Note that in case of the transform ${\cal L}$ the factors $f(j)$ of Eq.~(\ref{factor}) have to be taken properly into account.
In this way one obtains a set of functions $\tilde g_n(\sigma)$ which are then used for the expansion of the
transform, here given for the case of ${\cal L}$,
\begin{equation}
{\cal L}(\sigma) = \sum_{n=1}^N c_n \tilde g_n(\sigma) \,.
\end{equation} 
For given values of $N$ and
$\alpha$ of Eqs.~(\ref{inv1}) and (\ref{inv2}) a best fit to the calculated ${\cal L}$ is made, 
which determines the coefficients $c_n$. Varying then only the non-linear parameter $\alpha$
over a wide range values one obtains the absolute best fit for a specific $N$. Then one repeats the 
procedure increasing $N$ by one. A stable inversion result should be obtained in a range $N_A \le N \le N_B$.

In Fig.~7 we show inversion results of $L(\sigma)$ for the HH basis and of ${\cal L}$ for the Jacobi basis. 
The parameters for the HH basis are the same as defined in caption of Fig.~2d, for the Jacobi basis we use the new 
transform ${\cal L}$ with the setting
($N_1=24$, $N_2=79$) and ($N_1=5$, $N_2=19$) for $^3S_1$- and $^1S_0$-states in the pair coordinate, respectively. 
Note that for the HH basis we take $\sigma_I = 20$ MeV. We do not choose a higher resolution otherwise 
the inversion could be hampered too much by the fact that the low-energy strength is shifted to the peak region. 
Due to this misplaced strength one cannot expect that the two inversion results are extremely close to each other.
On the other hand, as Fig.~7a shows, differences remain rather small. The peak heights are almost identical,
but the peak of the HH basis is shifted somewhat to higher energies. 

It is a bit surprising that the low-energy cross 
sections are not completely different (see Fig.~7b), but this is due to the correct implementation of the Gamow factor
in the set of functions $g$ for the inversion. It is interesting to check the effect of an inversion of ${\cal L}$, 
where the Gamow factor in Eq.~(\ref{inv2}) is replaced by the factor $E^{3/2}$ 
(correct threshold behaviour without Coulomb barrier). Although we have a high-precision transform for the Jacobi basis
the inversion without the Gamow factor does not lead to the correct threshold behaviour, but at least coincides with the 
proper inversion result above about 6 MeV.
\vskip 1cm
\begin{figure}[htb]
\centerline{\includegraphics[width=0.65\columnwidth]{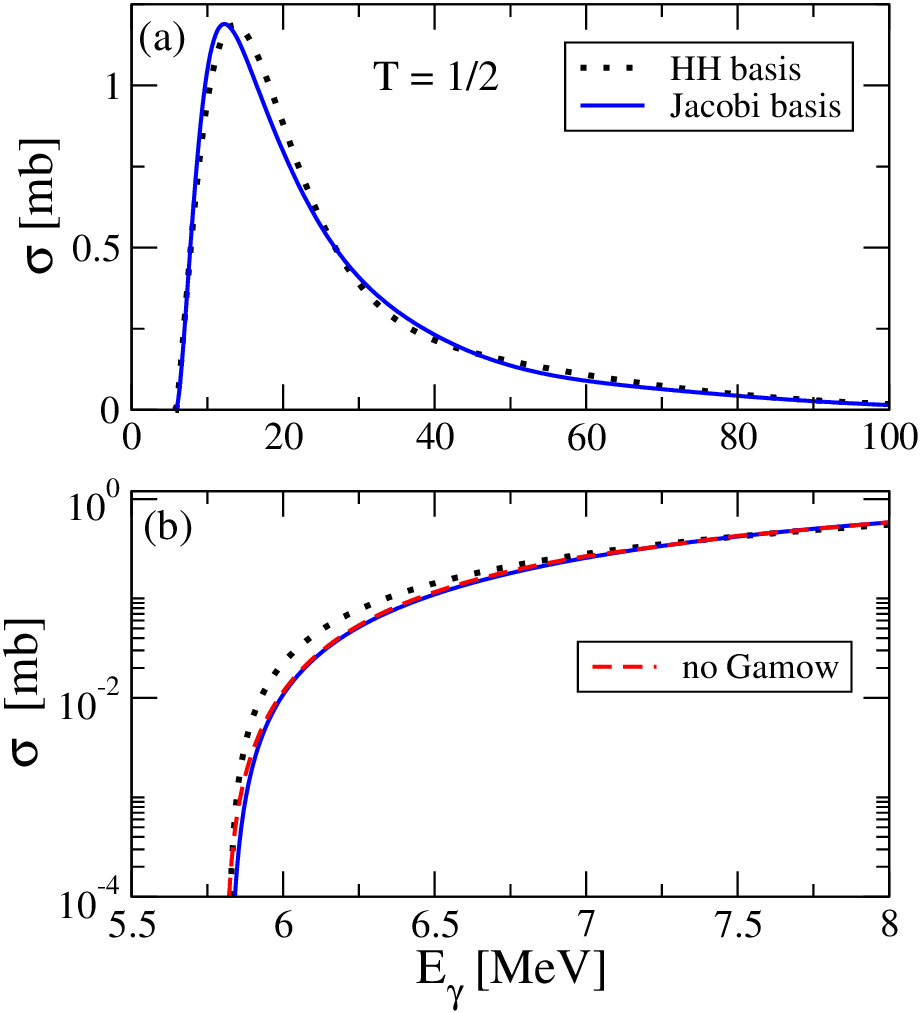}}
\caption{The $^3$He cross section $\sigma_\gamma$ of Eq.~(\ref{xsec}) obtained from inversions of $L(\sigma_R,\sigma_I=20$ MeV) 
with HH basis (dotted) and of ${\cal L}(\sigma)$ with Jacobi basis (full); in (b) also shown inversion of 
${\cal L}(\sigma)$ with Jacobi basis with factor $E^{3/2}$ instead of Gamow factor in functions $g$ of Eq.~(\ref{inv2})
(dashed).}
\end{figure}
In Fig.~8 we show a comparison of the LIT result with that of a calculation
with explicit continuum wave functions. In the upper panel the $^3$He
photodisintegration cross section $\sigma_\gamma$ is depicted. It is
evident that there is an excellent agreement between both results.
However, because of the strong fall-off of $\sigma_\gamma$ close
to the breakup threshold it is difficult to understand the level of 
agreement in this energy range. This can be estimated much better
for the $S$-factor since the Gamow factor is divided out. In Fig.~8b
one finds also in this case a very good agreement between both calculations.
It is worthwhile to mention that we find quite stable inversion results
$11 \le N \le 18$, where $N$ is the number of basis function used for the
inversion (see Eqs.~(\ref{inv1}) and (\ref{inv2})). This enables us
to make the following error estimate for the LIT inversion.
We take the inversions
for $N=11$ ($F_{\rm inv,11}(E)$) up to $N=18$ ($F_{\rm inv,18}(E)$)
and first determined an average inversion result
$\bar F_{\rm inv}(E) = \sum_{i=11}^{18}F_{\rm inv,i}(E)/8$, which is described by
the full cure in Fig.~8b. In addition we have calculated
the energy dependent standard deviation $\sigma_{\rm std}(E)$ and
the dashed curves correspond to $\bar F_{\rm inv}(E) \pm \sigma_{\rm std}(E)$.
As one sees the inversion error is rather small, but grows towards lower
energies. One could further improve the
inversions by making an even more precise LIT calculation. In our specific case
it would probably be better to change the $b_i$ parameters of the radial
basis function a bit rather than to increase the number of basis functions.
\vskip 1cm
\begin{figure}[htb]
\centerline{\includegraphics[width=0.65\columnwidth]{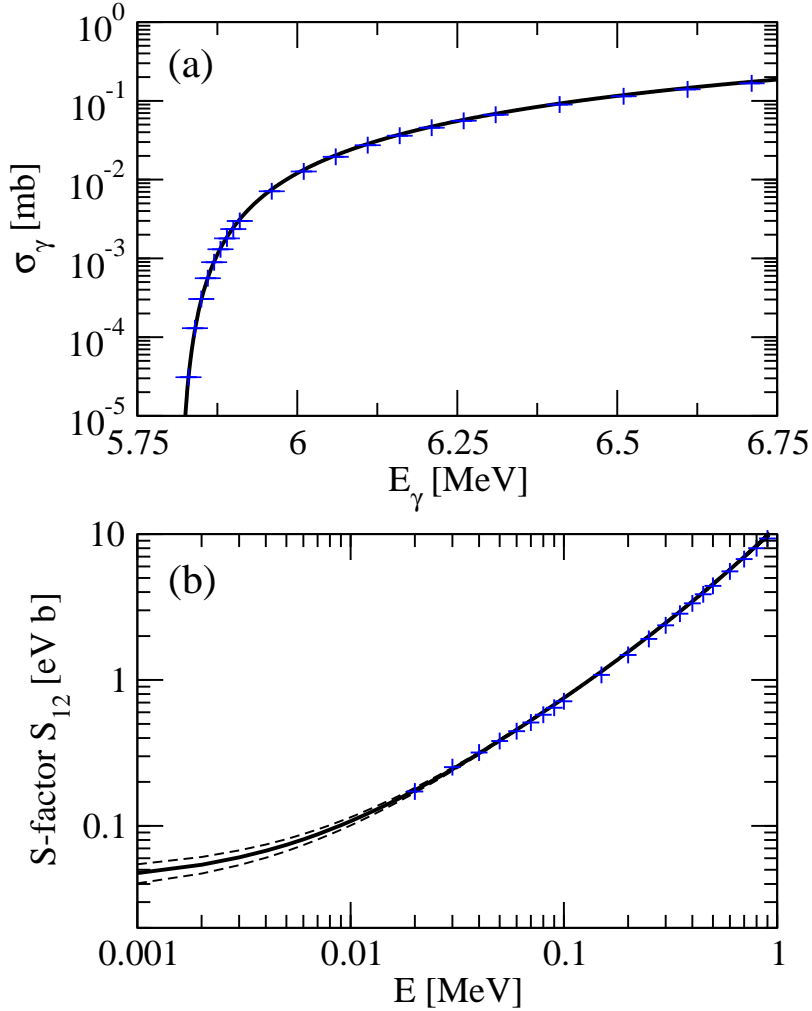}}
\caption{(a) Full curve same as in Fig.~7b and in addition results from the direct calculation with explicit continuum
wave function (plus signs); (b) same results as in (a) but rescaled in order to determine the $S$-factor (see Eqs. (\ref{S-fac})
and (\ref{cap})), inversion error shown by dashed lines (see text).
}
\end{figure}
We summarize our work as follows. We have tested the LIT method for a calculation of the $S$-factor of the 
reaction $^2$H$(p,\gamma)^3$He using a simple central NN interaction (MT-I/III potential). 
The calculation is performed by first computing the cross section
of the inverse reaction in unretarted dipole approximation and then using the law of detailed
balances in order to determine the deuteron-proton capture cross section which then
leads to the determination of the $S$-factor. For a precise application
of the LIT method it is necessary to have a sufficient density of LIT states in the energy region
of interest. Considering our specific case this corresponds for the $^3$He photodisintegration 
to the energy region between the two- and three-body breakup thresholds. We have found that a solution
of the LIT equation with the MT-I/III potential via an expansion in hyperspherical harmonics does not 
yield a single LIT-state below the three-body breakup threshold, even though using a rather high
number of basis functions of a rather large spatial extension. With a more realistic nuclear force the picture
does not change essentially as can be deduced from another low-energy observable, 
namely the $^4$He isoscalar
monopole resonance~\cite{BaB13,Lei15}. As pointed out in~\cite{Lei15}
for an increase of the LIT state density in the low-energy region
one needs to use a basis where the relevant dynamical variable, namely the single-particle coordinate
(vector pointing from the center of mass of the (A-1) particle system to the A-th particle), appears
explicitly. Therefore we have taken a basis which is a product of expansions of two basis systems, each of
them depending either on the single-particle coordinate or on the pair coordinate. We could show
that using such a basis one can systematically increase the low-energy LIT state density.
Furthermore, we show that in order 
to take into account that the LIT states become less dense with increasing energy it is advantageous 
to use different $\sigma_I$-values in different energy intervals.  

In addition to the LIT approach we have carried out the calculation with explicit contin-
uum wave functions. They have been determined via solving the Schr\"odinger equation with
the help of an expansion over a proper basis set. A comparison of results from both
methods shows a very good agreement. For the LIT method we have also included an estimate of the inversion error.

\end{document}